# AI Governance Control Stack for Operational Stability: Achieving Hardened Governance in AI Systems

Horatio Morgan

Morgan Signing House, AI Governance Architect

March 2026

**Abstract**

Artificial intelligence systems deployed in operational environments face increasing challenges related to governance, stability, and accountability. Traditional definitions of AI stability focus primarily on statistical performance or output reproducibility, but these definitions fail to capture the governance infrastructures required to maintain accountable AI operations over time. This paper proposes an AI Governance Control Stack, and a layered governance architecture designed to support operational stability through reproducible accountability artefacts including version governance, evidence-based verification, explainability logging, telemetry monitoring, drift detection, and escalation mechanisms. Drawing on research from machine learning operations, explainable AI, and regulatory governance frameworks, the paper introduces the concept of stability as the reproducibility of accountable system behaviour. The proposed architecture demonstrates how monitoring infrastructure, documentation artefacts, and institutional oversight mechanisms can be integrated into a unified governance framework capable of supporting stable and accountable enterprise AI deployment (NIST, 2023).

**Methodology**

The research adopted a conceptual architecture and pattern-based institutional analysis approach to examine how governance infrastructures can stabilize AI-enabled decision systems in enterprise environments. Rather than focusing solely on model performance metrics, the methodology integrates insights from machine learning operations research, explainable AI literature, and emerging AI governance frameworks to develop an operational control architecture. The research proceeds in three stages. First, existing literature on machine learning system reliability, concept drift, and explainable AI is analyzed to identify key operational risks associated with deploying AI systems in dynamic environments. This stage draws on research highlighting hidden technical debt in machine learning systems, the challenges of maintaining explainability in automated decision-making, and the need for continuous monitoring of deployed models. Second, a layered governance architecture is developed by synthesizing principles from software configuration management, regulatory compliance auditing, and AI accountability frameworks. The proposed architecture integrates four core governance mechanisms: system-of-record version governance, evidence-based control verification, decision-time explainability logging, and drift monitoring through structured stress testing. Third, the architecture is evaluated through pattern-based institutional analysis, examining how governance artefacts, telemetry monitoring systems, and escalation mechanisms interact to maintain stability in AI-enabled organizational environments. The analysis focuses on identifying governance patterns that enable institutions to detect

# AI Governance Control Stack for Operational Stability: Achieving Hardened Governance in AI Systems

behavioral change, preserve decision traceability, and enforce accountability across evolving AI deployments.

**Results**

The research identifies several governance mechanisms that are essential for maintaining operational stability in AI-enabled systems. First, system-of-record governance emerges as a critical stabilization mechanism. By binding model versions, datasets, policy configurations, and evaluation artefacts into a version-controlled governance bundle, organizations can prevent silent governance drift and ensure that system behaviour remains traceable across deployments. Second, evidence-based control verification enables organizations to demonstrate governance compliance through verifiable artefacts rather than declarative claims. The use of structured documentation frameworks, audit logs, and verification schedules provides a consistent mechanism for monitoring control effectiveness even as system behaviour evolves. Third, decision-time explainability logging provides a crucial accountability mechanism by capturing explanation artefacts at the moment of inference. These records allow governance teams to reconstruct decision contexts and identify behavioral shifts through mechanisms such as explanation delta analysis. Fourth, multi-dimensional drift monitoring includes semantic, behavioral, and probabilistic drift signals provide a comprehensive approach to detecting changes in system behaviour that may not be visible through traditional performance metrics. When combined with telemetry monitoring and replicable stress-testing environments, these mechanisms enable organizations to identify anomalies and trigger governance escalation procedures. Collectively, these findings demonstrate that AI stability depends on the integration of governance artefacts, explainability infrastructure, and monitoring systems into a unified control architecture.

**Conclusion**

This study proposes an AI Governance Control Stack for Operational Stability, a layered architecture designed to maintain traceability, accountability, and resilience in AI-enabled decision systems. The research demonstrates that stability in AI systems cannot be achieved solely through improvements in model performance or robustness. Instead, stability emerges from governance infrastructures capable of detecting behavioral change, maintaining decision traceability, and enforcing institutional oversight when risk thresholds are exceeded. By integrating system-of-record governance, evidence-based control verification, decision-time explainability logging, and multi-dimensional drift monitoring, the proposed framework provides a practical approach to operationalizing responsible AI principles within enterprise environments. The findings also highlight the importance of telemetry monitoring, gated propagation controls, and governance escalation mechanisms in ensuring that AI autonomy remains subject to human authority. As AI systems continue to increase in capability and operational impact, governance architectures must evolve to ensure that accountability remains embedded within technical infrastructures. Future research should therefore explore empirical validation of governance control stacks across

# AI Governance Control Stack for Operational Stability: Achieving Hardened Governance in AI Systems

organizational contexts and examine how institutional governance mechanisms can adapt to increasingly autonomous AI systems.

## 1. Introduction

Artificial intelligence systems increasingly operate as components of enterprise decision infrastructures rather than isolated analytical tools. From financial services and healthcare to logistics and public administration, organizations are integrating AI models into automated workflows that directly influence operational and strategic decisions. As these systems scale in capability and autonomy, governance frameworks must ensure that AI behaviour remains stable, traceable, and auditable over time.

However, maintaining operational stability in AI systems presents significant challenges. Machine-learning systems deployed in production environments often experience instability due to data distribution shifts, evolving operational contexts, and undocumented system changes. Research on machine-learning infrastructure has shown that hidden technical debt arises from untracked dependencies, dataset evolution, and configuration drift can undermine reliability even when model accuracy appears stable (Sculley et al., 2015). Similarly, empirical studies of machine-learning engineering practices highlight the complexity of maintaining AI systems across the full lifecycle of development, deployment, monitoring, and iteration (Amershi et al., 2019). These challenges are particularly pronounced in generative AI systems, whose outputs depend not only on model parameters but also on prompts, retrieval data, and runtime context. As a result, behavioral changes may occur even when underlying models remain unchanged. Traditional evaluation metrics such as accuracy or perplexity are therefore insufficient indicators of governance stability.

Regulatory frameworks are increasingly recognizing this issue. The European Union (EU) Artificial Intelligence Act requires high-risk AI systems to maintain technical documentation, automatic logging, human oversight, and post-deployment monitoring mechanisms (European Commission, 2024). Similarly, the NIST AI Risk Management Framework (AI RMF) emphasizes lifecycle governance through monitoring, incident response, and continuous improvement processes (NIST, 2023). These developments suggest that AI stability must be understood not simply as technical performance persistence but as the organization's ability to demonstrate reproducible accountability across the AI lifecycle. Focus is therefore placed on a governance control stack architecture designed to operationalize AI stability through integrated governance and monitoring mechanisms. The framework combines version governance, evidence-based verification, explainability logging, telemetry monitoring, and drift detection to enable organizations to maintain consistent oversight of AI behaviour across system updates and operational contexts. This paper research proposes an AI Governance Control Stack for

# AI Governance Control Stack for Operational Stability: Achieving Hardened Governance in AI Systems

Operational Stability, which integrates system-of-record governance, evidence-based control verification, decision-time explainability logging, and drift monitoring mechanisms.

## 2. Related Work

Research on trustworthy AI has increasingly emphasized the importance of governance mechanisms that enable transparency, reproducibility, and accountability in machine learning systems. Documentation frameworks such as Model Cards provide structured reporting mechanisms for model behaviour and limitations (Mitchell et al., 2019). Lifecycle management frameworks such as MLflow enable version governance and experiment tracking that support reproducibility and auditing of machine learning models (Zaharia et al., 2018).

Operational monitoring has also emerged as a critical requirement for production machine learning systems. Production readiness frameworks emphasize structured validation and testing processes that detect performance degradation and technical debt before models affect real-world decisions (Breck et al., 2017). Additionally, production-scale machine learning platforms integrate telemetry monitoring and pipeline validation to ensure reliable model operation (Baylor et al., 2017). Model stability is also affected by distribution shift between training and operational environments, requiring continuous drift detection and monitoring mechanisms (Lipton, Wang and Smola, 2018). Governance frameworks further emphasize the need for internal auditing and escalation mechanisms that allow organizations to investigate failures in automated decision systems (Raji et al., 2020).

Building on these contributions, this paper proposes an integrated governance architecture that connects monitoring infrastructure, explainability artefacts, and accountability processes into a unified operational governance framework.

## 3. Governance Stability Problems

Stability in operational AI systems cannot be understood solely in terms of statistical model performance. Instead, stability must also incorporate governance capabilities that allow decisions to be reproduced, audited, and verified across time. In this context, stability can be understood as the reproducibility of accountable system behaviour, supported by governance artefacts such as documentation, telemetry logs, and version-controlled model deployments. This study adopts a conceptual design-science methodology, which focuses on developing governance architectures grounded in theoretical and practical evidence. Rather than empirically evaluating a specific deployed system, the paper synthesizes insights from three bodies of literature. First, research on machine-learning operations and software engineering provides evidence on the operational challenges associated with deploying AI systems at scale, including hidden technical debt, infrastructure coupling, and lifecycle instability (Sculley et al., 2015; Amershi et al., 2019). Second, scholarship on explainability and algorithmic accountability provides theoretical foundations for transparency, interpretability, and institutional oversight in automated decision

# AI Governance Control Stack for Operational Stability: Achieving Hardened Governance in AI Systems

systems (Doshi-Velez and Kim, 2017; Mittelstadt et al., 2016). Third, the paper draws on sector and regulatory governance frameworks, including the NIST AI RMF, the EU AI Act, and ISO/IEC 42001, which establish lifecycle requirements for AI monitoring, documentation, risk management, and human oversight. Hence, by synthesizing these sources, the paper constructs a governance architecture that is: normatively grounded in responsible AI principles, operationally plausible within enterprise AI environments and aligned with emerging regulatory expectations for trustworthy AI deployment.

## 4. Literature Review

### 4.1 Explainability and Accountability in AI Systems

Explainability has become a central concept in responsible AI governance. Interpretable machine-learning approaches seek to provide insights into how models generate predictions or decisions. Doshi-Velez and Kim (2017) opines that interpretable models must be evaluated rigorously to ensure that explanations genuinely reflect model behaviour rather than superficial approximations. However, scholars emphasize that explainability alone does not guarantee accountability. Mittelstadt et al. (2016) argue that algorithmic transparency must be embedded within institutional governance structures that enable scrutiny, contestability, and responsibility for automated decisions. Without such institutional mechanisms, explainability risks becomes a symbolic risk rather than operational feature of AI governance.

### 4.2 Operational Instability in Production AI Systems

Research on production machine-learning systems highlights the difficulty of maintaining stable AI behaviour over time. It was noted that machine-learning infrastructure often accumulates hidden technical debt through undocumented dependencies, evolving datasets, and configuration changes that alter system behaviour (Sculley et al., 2015). Similarly, it was interesting to realize that the lifecycle management of machine-learning systems involves complex interactions between data engineering, software infrastructure, and organizational processes (Amershi et al., 2019). These findings suggest that stability in AI systems cannot be achieved through model design alone but require comprehensive governance mechanisms spanning the entire system lifecycle. In support, Raji et al., (2020) argued that institutional governance structures must incorporate continuous auditing processes that evaluate model behavior across the entire deployment lifecycle.

### 4.3 Governance Frameworks and Lifecycle Monitoring

Regulatory frameworks increasingly emphasize lifecycle governance in AI systems. The NIST AI Risk Management Framework highlights monitoring, incident response, and continuous improvement as core components of trustworthy AI deployment (NIST, 2023). The EU AI Act similarly mandates risk management processes, technical documentation, logging, and human oversight mechanisms for high-risk AI systems (European Commission, 2024). International standards also reflect this shift. ISO/IEC 42001 establishes a management-system approach to AI governance, emphasizing organizational accountability, risk management, and continual improvement in AI lifecycle management.

### 4.4 Emerging Perspectives on AI Resilience and Capability-Sensitive Governance

Recent developments in AI safety research suggest that governance frameworks must increase scale with system capabilities rather than simply intended use cases. Frontier research initiatives such as the Laboratory for AI Security Research (LASR) led by the University of Oxford in collaboration with the Alan Turing Institute and UK government partners focus on understanding how advanced AI systems behave under adversarial or high-impact conditions (University of Oxford, 2024; The Alan Turing Institute, 2024; UK Department for Science, Innovation and Technology, 2024). This work highlights the importance of AI resilience, defined not only as robustness against technical failures but as the ability of governance systems to maintain oversight and control under stress conditions. Stress testing therefore becomes a form of governance resilience testing, evaluating whether escalation mechanisms trigger appropriately, whether human authority intervenes when risk thresholds are exceeded, and whether systems degrade safely under uncertainty or misuse. These developments reinforce the need for hardened governance architectures capable of maintaining accountability as AI capabilities continue to evolve.

### 4.5 Stability as Reproducibility of Accountability

Traditional definitions of AI stability focus on reproducibility of outputs or persistence of performance metrics. However, such definitions are insufficient in enterprise contexts where AI systems operate within complex socio-technical infrastructures. In this paper, stability is defined as the reproducibility of accountability. A stable AI system is one in which organizations can reproduce the governance context under which a decision was generated. This includes: the model version used, the datasets and retrieval sources involved, the governance policies in effect, the explanation artefacts produced, the monitoring signals and escalation outcomes recorded. This definition aligns technical reliability with institutional accountability and provides a practical foundation for operational governance. In addition, Doshi-Velez and Kim (2017) highlighted how the application of the Explainable AI approach aims to provide interpretable representations of model decisions that support accountability and reproducibility in automated decision systems; while Leslie, (2019) focused on understanding artificial intelligence ethics and safety as a guide for the responsible design and implementation of AI systems in the public sector.

### 5. Research Contribution to the Advancement of governance control stack architecture

This paper contributes to the literature by proposing a governance control stack architecture that integrates these previously fragmented domains into a unified operational framework for maintaining stability in explainable AI systems. First, it introduces the concept of stability as reproducibility of accountability, reframing stability in AI systems as the ability to reproduce governance context including model versions, datasets, policies, and explanation artefacts rather

# AI Governance Control Stack for Operational Stability: Achieving Hardened Governance in AI Systems

than merely reproducing technical outputs (European Commission, 2024). This conceptual reframing connects explainability research with institutional governance practices and provides a practical definition of stability suitable for enterprise AI deployment. Second, the paper proposes a layered governance control stack linking version governance, evidence-based verification, decision-time explainability logging, telemetry monitoring, drift detection, and governance escalation mechanisms (Amodei et al., 2016). By integrating these components, the framework translates high-level responsible AI principles into operational processes that can be implemented within AI lifecycle management systems. Third, the paper extends emerging work on AI resilience by introducing a capability-aware hardened governance perspective, arguing that governance intensity must scale with AI system capability and operational impact. In this model, stress testing becomes a form of governance resilience testing, evaluating whether escalation pathways, human oversight mechanisms, and accountability structures remain effective when AI systems operate under adverse or high-risk conditions (OpenAI, 2025).

Together, these contributions bridge the gap between technical explainability research, machine-learning operations practice, and institutional AI governance frameworks. The resulting architecture provides a foundation for translating responsible AI principles into repeatable, auditable governance infrastructures capable of supporting large-scale enterprise AI deployment.

Figure 1. Conceptual framework linking explainability, accountability, governance infrastructure, and resilience in achieving operational AI stability.

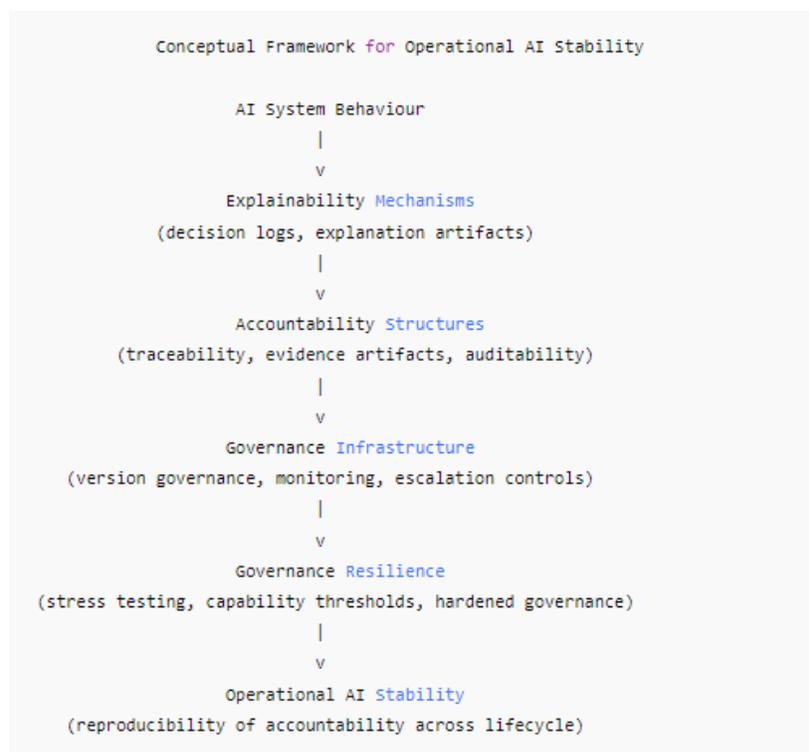

# AI Governance Control Stack for Operational Stability: Achieving Hardened Governance in AI Systems

The framework illustrates how explainability mechanisms support accountability structures, which in turn depend on governance infrastructures capable of monitoring system behaviour and enforcing escalation responses. Resilience mechanisms ensure that these governance structures remain effective under stress conditions.

The conceptual framework illustrates the theoretical foundation of the governance control stack proposed in this paper. Explainability mechanisms provide the first layer of transparency by generating interpretable records of system behaviour. However, explanations alone do not guarantee accountability. Accountability emerges when explanation artefacts are embedded within traceable evidence structures that allow decision processes to be reviewed and audited (European Commission, 2024). These structures must be supported by governance infrastructures capable of monitoring system behaviour, maintaining version-controlled artefacts, and enforcing escalation mechanisms when risk thresholds are exceeded (OpenAI, 2025). Finally, governance resilience mechanisms—such as stress testing and capability-aware control thresholds—ensure that accountability structures remain effective when AI systems operate under adverse or high-risk conditions. Operational stability therefore emerges from the interaction between explainability, accountability, governance infrastructure, and resilience mechanisms.

## 6. AI Governance Control Stack for Architecture

### 6.1 System-of-Record Version Governance

The first layer of the control stack is a canonical system-of-record governance artefact. This artefact combines model versions, dataset identifiers, configuration parameters, and governance policies into a traceable bundle maintained through semantic versioning. Such version governance ensures that every stress-test run and operational decision can be tied to a specific governance context. This approach mirrors configuration management practices used in large-scale software systems and helps prevent silent governance drift.

# AI Governance Control Stack for Operational Stability: Achieving Hardened Governance in AI Systems

Figure 2. Governance Control Stack for AI Stability

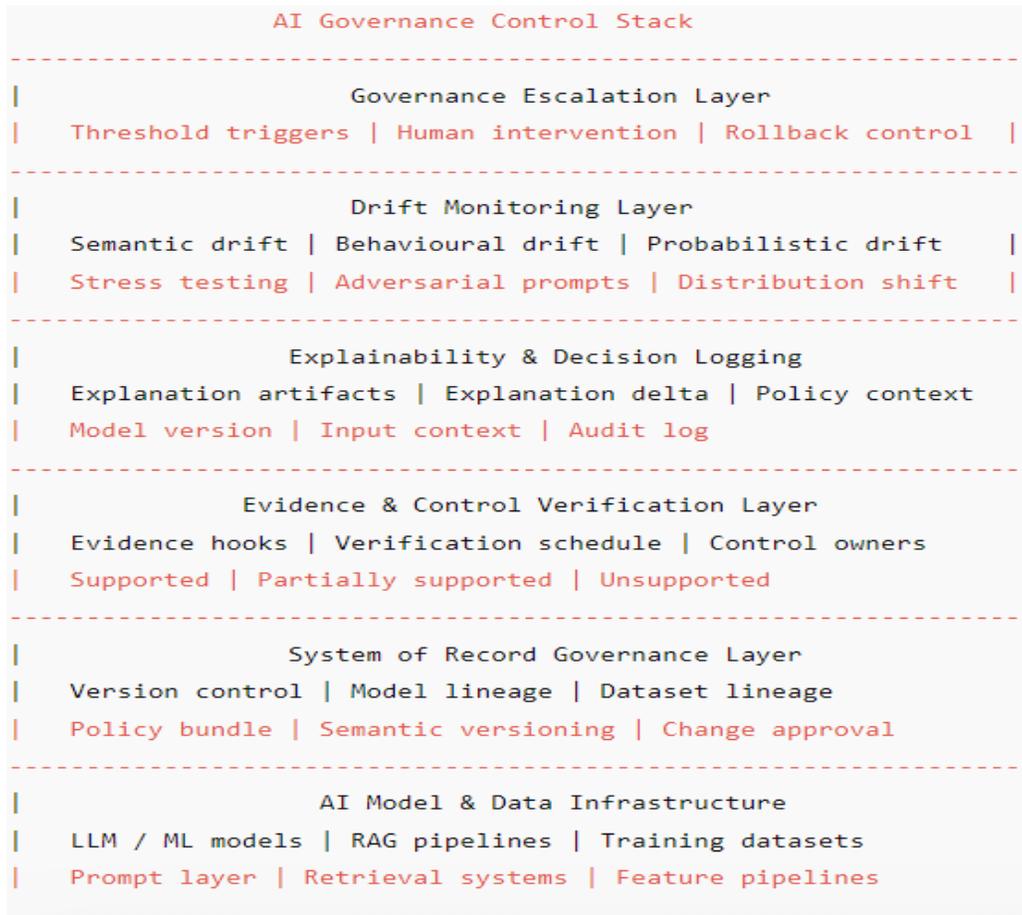

The control stack presented in Figure 2 operationalizes the central argument of this paper: that stability in AI systems must be achieved through governance infrastructures capable of maintaining traceability, accountability, and controlled autonomy over time. The layered control stack is used to maintain stability, traceability, and governance oversight in AI-enabled decision systems. The architecture integrates technical monitoring mechanisms with institutional governance controls, forming a multi-layer framework that connects model infrastructure with accountability processes. At the base of the architecture are the AI models, datasets, and retrieval pipelines that generate system outputs. Above this layer, system-of-record governance establishes version control, dataset lineage, and policy configuration management, ensuring that system changes remain traceable across deployments. Evidence verification and explainability logging layers provide documentation artefacts and decision-time transparency, allowing governance teams to reconstruct the reasoning context associated with automated outputs. Drift monitoring mechanisms then evaluate semantic, behavioral, and probabilistic changes in system behaviour through structured stress-testing and telemetry signals. Finally, a governance escalation layer enables institutional oversight by defining thresholds that trigger investigation, rollback, or human intervention when anomalies are detected. Together, these layers form an operational governance architecture that ensures stability not only in model performance but also in organizational accountability for AI-enabled decisions.

**6.2 Evidence Hooks and Atomic Control Verification**

Governance controls in AI systems must generate verifiable artefacts rather than rely on declarative compliance statements. In operational environments, accountability depends on the ability of organizations to demonstrate that governance controls are functioning as intended at specific points in time. Each government control should therefore include clearly defined evidence hooks that specify the artefacts required to demonstrate control effectiveness. These artefacts may include system logs, configuration records, model documentation, evaluation reports, and audit trails. In addition to identifying required artefacts, each control must specify verification procedures, accountable control owners, and verification schedules that determine how frequently compliance is assessed. To ensure consistent evaluation, controls can be assessed using atomic verification states, such as supported, partially supported and unsupported

This atomic verification approach mirrors audit methodologies used in regulatory compliance frameworks, where control assertions must be supported by traceable evidence rather than informal assurances. Evidence artefacts may include documentation such as model cards, which provide structured reporting of model characteristics, intended uses, and evaluation results (Mitchell et al., 2019), as well as dataset documentation frameworks that describe training data provenance, collection methods, and limitations (Gebru et al., 2021). By structuring governance controls around explicit evidence artefacts, organizations can maintain consistent oversight even as models, datasets, and operational environments evolve. Evidence-based verification also supports external auditability and regulatory compliance, particularly under governance frameworks that emphasize documentation, traceability, and lifecycle accountability.

**7. Alignment with Existing AI Governance Frameworks**

7.1 Alignment with other frameworks

The AI Governance Control Stack proposed in this paper aligns with and extends existing AI governance frameworks that emphasize monitoring, accountability, and lifecycle oversight. Recent regulatory and policy initiatives—including the NIST AI Risk Management Framework, the European Union Artificial Intelligence Act, and the ISO/IEC 42001 AI management system standard—highlight the importance of documentation, monitoring, transparency, and human oversight in the governance of AI systems. However, these frameworks primarily articulate governance principles and organizational requirements rather than specifying operational architectures capable of implementing these principles within deployed AI systems. The

# AI Governance Control Stack for Operational Stability: Achieving Hardened Governance in AI Systems

governance control stack introduced in this paper translates these governance requirements into operational infrastructure layers capable of supporting stable and accountable AI deployment.

Table 1 illustrates how the layers of the AI Governance Control Stack correspond to governance requirements found in major AI governance frameworks.

| Governance Layer | Governance Function | Corresponding Framework Requirements |
|---|---|---|
| System-of-Record Version Governance | Traceable model and policy versioning | NIST AI RMF – Governance and Documentation |
| Evidence-Based Verification | Verification of governance controls and evidence artefacts | ISO/IEC 42001 – AI Management Systems |
| Decision-Time Explainability Logging | Documentation of decision reasoning | EU AI Act – Transparency and Logging |
| Telemetry Monitoring | Continuous monitoring of model behaviour | NIST AI RMF – Monitoring and Risk Management |
| Drift Detection | Detection of statistical and behavioral instability | AI Safety and ML Monitoring Research |
| Governance Escalation | Human oversight and incident response | EU AI Act – Human Oversight Requirements |

The comparison demonstrates that the governance control stack operationalizes governance requirements that are often expressed only at a policy or regulatory level. By embedding monitoring, documentation, and escalation mechanisms directly within AI system infrastructures, the proposed architecture enables organizations to translate governance principles into operational practice.

## 8. Runtime Governance and Behavioral Stability Monitoring

### 8.1 Decision-Time Explainability Logging

Explainability mechanisms provide an additional stabilization layer within the governance control stack. Rather than generating explanations only during retrospective audits, explanation artefacts should be captured at the moment of inference, ensuring that decision context is preserved before

# AI Governance Control Stack for Operational Stability: Achieving Hardened Governance in AI Systems

system state changes occur (OpenAI, 2025). Decision-time explainability logs typically include model version identifier, input context or prompt, explanation artefact (e.g., feature attribution or reasoning trace) and policy bundle or governance configuration version. Capturing explanation artefacts at decision time ensures that organizations retain an immutable record of how automated decisions were produced within a specific governance context. These records complement model documentation practices such as model cards and dataset documentation—which are designed to improve transparency and accountability in machine learning systems (Mitchell et al., 2019; Gebru et al., 2021).

An additional mechanism, referred to in this paper as explanation delta, tracks how explanation artefacts change across model versions or system updates. By comparing explanation patterns over time, governance teams can identify subtle behavioral shifts that may not be visible through traditional performance metrics alone. Explanation deltas therefore provide an early indicator of potential governance drift, enabling organizations to distinguish legitimate model improvements from unintended behavioral deviations.

8.2 Telemetry, Monitoring and Governance

Operational stability in AI systems requires continuous monitoring of system behaviour after deployment. As a result of which continuous telemetry and model observability have become core requirements for operational AI stability, enabling real-time monitoring of prediction drift, performance degradation, and data pipeline anomalies (Datadog, 2023). Telemetry monitoring provides the primary mechanism for observing system performance and identifying anomalies that may indicate drift, misuse, or system degradation. Telemetry signals may include inference behaviour patterns, refusal or safety-trigger rates, latency anomalies, policy violations, retrieval failures in Retrieval Augmented Generation (RAG) pipelines and unexpected output distributions. However, telemetry monitoring alone is insufficient unless it is embedded within organizational governance processes. In enterprise environments, telemetry alerts must be linked to named governance owners and escalation pathways, ensuring that anomalies trigger investigation and corrective action rather than remaining passive system signals. A complementary mechanism for maintaining stability is gated propagation, which prevents unverified system changes from reaching production environments (Breck, E. et al. 2021). Under gated propagation, updates to models, prompts, retrieval pipelines, or policy bundles must pass predefined validation checks including evaluation tests, governance approval, and monitoring readiness before deployment is permitted.

Together, telemetry monitoring and gated propagation enable what this paper describes as hardened governance (NIST, 2023). Hardened governance represents a shift from reactive response to where governance mechanisms respond to failures after they occur toward proactive control of system autonomy. In this model, governance infrastructure continuously monitors system behaviour, enforces change-management discipline, and ensures that escalation mechanisms activate when risk thresholds are exceeded. As AI systems increase in capability and operational impact, such hardened governance architectures become essential for preserving institutional accountability and ensuring that human oversight remains effective. In this sense, operational stability emerges not from static model performance but from governance architectures capable of detecting behavioral change, enforcing accountability, and ensuring that system autonomy remains subject to institutional control (ISO/IEC 42001).

8.3 Drift Monitoring and Resilience Testing

Drift monitoring in generative AI systems must account for multiple dimensions of change. Semantic drift refers to changes in the meaning or interpretation of generated outputs. Behavioral drift occurs when models deviate from governance policies or safety constraints. Probabilistic drift reflects changes in output distribution patterns across repeated inference runs. Distribution shift between training and production environments remains one of the primary causes of model instability, requiring detection mechanisms that monitor changes in data distribution and prediction behavior (Lipton et al., 2018). Hence, monitoring all three dimensions is necessary because generative systems may produce semantically divergent outputs while maintaining stable statistical performance metrics. Replicable resilience-testing environments using golden prompt sets, adversarial scenarios, pinned datasets, and fixed evaluation rubrics enable organizations to detect such drift reliably.

# AI Governance Control Stack for Operational Stability: Achieving Hardened Governance in AI Systems

Figure 3. AI Stability Monitoring Lifecycle.

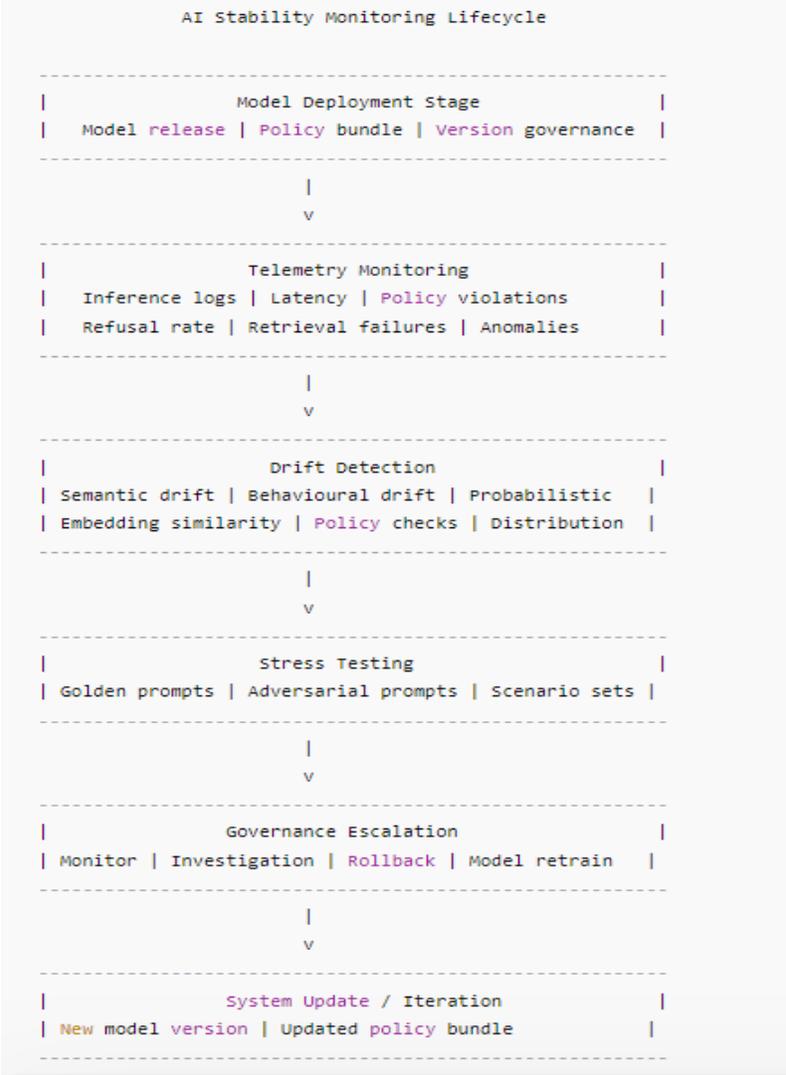

The lifecycle illustrates how monitoring, stress testing, and governance escalation interact to maintain stability in AI systems. Telemetry signals trigger drift detection mechanisms, which may initiate stress testing and governance interventions before model updates are propagated.

Maintaining stability in AI systems requires continuous lifecycle monitoring rather than one-time evaluation. The stability monitoring lifecycle illustrates how telemetry, drift detection, and governance escalation interact within a continuous feedback loop. This is applicable in a space where production machine learning systems require rigorous operational testing frameworks to detect degradation and technical debt before models affect real-world decisions (Breck et al., 2017). This approach is recognizable in its unique importance to continuous lifecycle monitoring traceable AI deployment pipelines (Zaharia, M., et al., 2018).

# AI Governance Control Stack for Operational Stability: Achieving Hardened Governance in AI Systems

After deployment, telemetry signals capture inference behaviour and system anomalies. Drift detection mechanisms analyze these signals to identify semantic, behavioral, or probabilistic deviations from baseline behaviour. Stress-testing environments then evaluate whether detected changes represent legitimate model improvements or governance risks. If risk thresholds are exceeded, governance escalation mechanisms trigger interventions such as investigation, rollback, or model retraining. Updated models and policy bundles are subsequently redeployed, completing the governance feedback cycle.

## 9. Enterprise and Sector Implications

Enterprise AI transformation literature emphasizes that successful AI deployment depends on governance, data quality, organizational readiness, and monitoring capabilities rather than isolated pilot projects. Many organizations remain in early experimentation phases because governance mechanisms have not matured sufficiently to support large-scale deployment. The governance control stack proposed in this paper is particularly relevant in regulated sectors such as finance, healthcare, and public administration, where documentation, logging, human oversight, and incident response are mandatory requirements. In these environments, explainability must function as operational governance infrastructure rather than merely a diagnostic feature (NIST, 2023).

The relationship between AI capability and governance intensity can be conceptualized as a capability-aware control model, where higher system capabilities require progressively stronger governance and resilience mechanisms.

# AI Governance Control Stack for Operational Stability: Achieving Hardened Governance in AI Systems

Figure 4. Capability-Aware Hardened Governance Model.

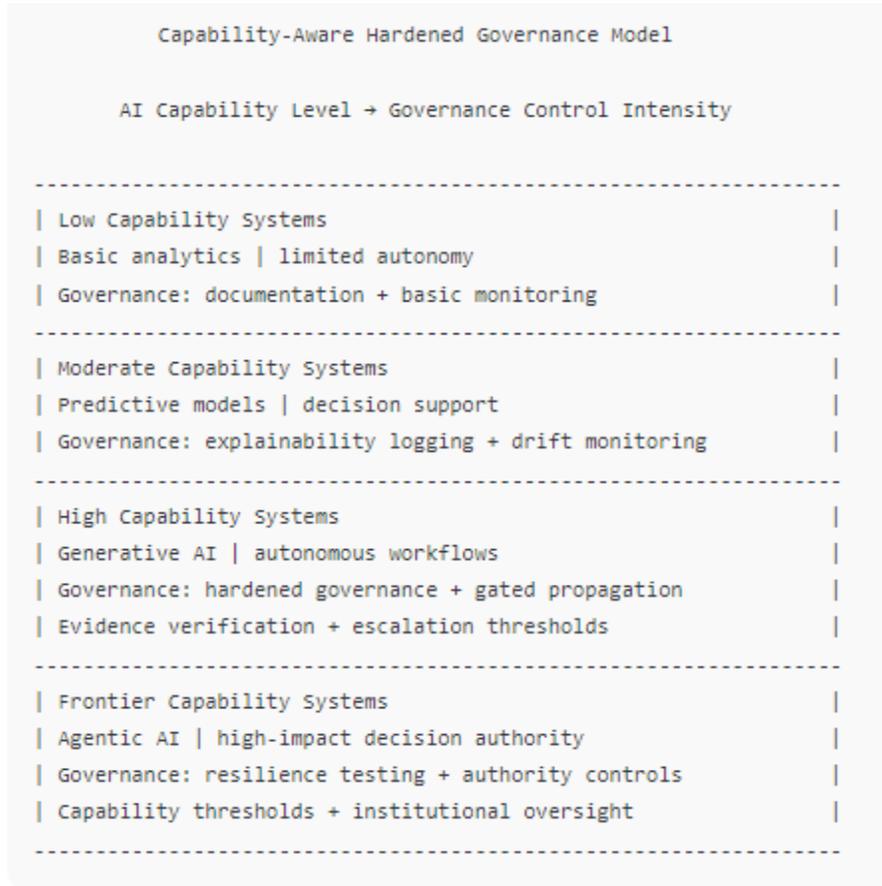

The model illustrates how governance intensity must increase as AI system capabilities expand. Higher capability systems require stronger oversight mechanisms, including resilience testing, authority controls, and capability-sensitive governance thresholds.

The capability-aware governance model illustrates how governance intensity must scale with the capabilities and potential impact of AI systems. Low-capability analytical systems may require only documentation and basic monitoring mechanisms. However, as AI systems become capable of autonomous decision-making or generative reasoning, governance mechanisms must expand to include explainability logging, drift monitoring, and formal evidence verification processes. At higher capability levels—such as systems supporting autonomous workflows or agentic behaviour—governance must become hardened, incorporating gated propagation, escalation thresholds, and explicit human authority controls. For frontier or high-impact systems, governance resilience becomes critical, requiring stress testing, institutional oversight, and capability-sensitive control mechanisms designed to ensure that human authority remains effective even when systems operate under adversarial or uncertain conditions (Miller, 2021).

**10. Discussions**

# AI Governance Control Stack for Operational Stability: Achieving Hardened Governance in AI Systems

The governance control stack proposed in this paper contributes to an emerging shift in AI governance research from static compliance frameworks toward dynamic lifecycle governance models. Much of the existing literature on responsible AI focuses on normative principles such as fairness, transparency, and accountability (Mittelstadt et al., 2016). While these principles remain essential, scholars increasingly recognize that the operationalization of these values within deployed AI systems remains an unresolved challenge. The architecture proposed here suggests that explainability should be conceptualized not as an isolated technical feature but as a component of a broader governance infrastructure that links monitoring, evidence generation, and escalation mechanisms. This perspective aligns with recent research emphasizing that trustworthy AI requires integrated socio-technical systems combining technical safeguards with organizational governance structures (NIST, 2023).

A second implication concerns the growing importance of AI resilience and capability-sensitive governance. As AI systems become more capable and embedded in critical decision processes, governance models must account for the interaction between system capability, operational risk, and institutional control. The concept of hardened governance introduced in this paper suggests that governance intensity should be scaled with the capabilities and potential impact of AI systems. In this framework, stress testing becomes a form of governance resilience evaluation, assessing whether monitoring systems, escalation pathways, and human oversight mechanisms remain effective under adversarial or high-risk conditions. This approach complements emerging research in AI safety and resilience that emphasizes the importance of robustness, reliability, and accountability as interdependent dimensions of trustworthy AI systems (Amodei et al., 2016).

Finally, the framework highlights the need for future empirical research on operational AI governance infrastructures. While conceptual frameworks and regulatory guidelines are expanding rapidly, comparatively little empirical work examines how organizations implement monitoring, explainability logging, or governance escalation mechanisms in real-world AI deployments. Future research could therefore investigate how governance control stacks function across sectors such as finance, healthcare, and public administration, and evaluate how telemetry monitoring and replicability is. This paper argues that stability in AI systems should be understood as the reproducibility of accountability rather than merely the reproducibility of outputs (Pineau et al., 2021). By integrating explainability mechanisms with governance monitoring, version control, and escalation pathways, the proposed governance control stack provides an operational architecture for maintaining stability across the AI lifecycle. resilience testing, and capability thresholds influence organizational oversight and risk management practices (NIST 2023).

Limitations and Non-Claims

This paper does not claim that the proposed governance of architecture eliminates all forms of AI risk or drift. Rather, it proposes a framework for improving organizational capacity to detect and manage behavioral change in AI systems. The architecture is conceptual and requires empirical validation in real-world enterprise deployments. Future research should examine how governance control stacks perform in specific sectors and evaluate the trade-offs between governance depth, operational cost, and deployment agility.

## 11. Conclusion

This paper argues that stability in AI systems should be understood as the reproducibility of accountability rather than merely the reproducibility of outputs. By integrating explainability mechanisms with governance monitoring, version control, and escalation pathways, the proposed governance control stack provides an operational architecture for maintaining stability across the AI lifecycle. Maintaining stability in AI systems requires more than monitoring model performance. It requires governance infrastructures capable of detecting behavioral change, preserving decision authority, and enforcing accountability responses. This paper proposes a layered governance architecture integrating version governance, evidence-based verification, explainability logging, telemetry monitoring, drift detection, and escalation mechanisms (Zaharia et al., 2018). Together, these elements enable organizations to maintain reproducible accountability across the AI lifecycle. As AI systems continue to increase in capability, the challenge of governance will shift from compliance documentation toward resilience under stress, where the durability of oversight mechanisms becomes the defining characteristic of trustworthy AI deployment. As AI capabilities increase, governance can no longer rely on static compliance frameworks alone. Instead, stability will depend on governance architectures that combine explainability, monitoring, resilience testing, and capability-aware oversight to ensure that human authority remains effective in increasingly autonomous systems (Mitchell et al., 2019).

# AI Governance Control Stack for Operational Stability: Achieving Hardened Governance in AI Systems

# AI Governance Control Stack for Operational Stability: Achieving Hardened Governance in AI Systems